\documentclass[namedreferences]{solarphysics}
\usepackage{spr-sola-addons} 
\usepackage{color}           
\usepackage{url}             

\usepackage{graphicx} 

\newcommand{\arcmin}{{\hbox{ $^\prime$}}}
\newcommand{\arcsec}{{\hbox{$^{\prime\prime}$}}}

\newcommand{\be}{\begin{equation}}
\newcommand{\ee}{\end{equation}}
\newcommand{\bea}{\begin{eqnarray}}
\newcommand{\eea}{\end{eqnarray}}
\newcommand{\beas}{\begin{eqnarray*}}
\newcommand{\eeas}{\end{eqnarray*}}

\newcommand{\pasj}{   {\it Pub. Astron. Soc. Japan}}

\begin{document}

\begin{article}

\begin{opening}

\title{RATAN-600 Observations of Small Scale Structures with High Spectral Resolution
}

\author{V. M.~\surname{Bogod}$^{1,2}$\sep
        C. E.~\surname{Alissandrakis}$^{3}$\sep
        T. I.~\surname{Kaltman}$^{1}$\sep      
        S. Kh.~\surname{Tokhchukova}$^{1}$      
       }
\runningauthor{Bogod et al.}
\runningtitle{RATAN-600 Observations of Small Scale Structures}

   \institute{
$^{1}$ Special Astrophysical Observatory, Russia
email: \url{vbog_spb@mail.ru} \\ 
$^{2}$ St Petersburg National Research University ITMO, Russia\\
$^{3}$ Section of Astro-Geophysics, Department of Physics, University 
of Ioannina, GR-45110 Ioannina, Greece  
email: \url{calissan@cc.uoi.gr} \\
             }

\begin{abstract}
We present observations of quiet-sun small-scale structures (SSS) in the microwave range with the {\it Radio Astronomical Telescope of the Academy of Sciences 600} (RATAN-600) spectral-polarization facility in a wide range of frequencies. SSS are regularly recorded in routine observations of the large reflector-type radio telescope and represent manifestations in the radio range of various structures of the quiet-sun: supergranulation network, bright points, plage patches and so on. A comparison with with images from the {\it Solar Dynamics Observatory} (SDO) showed that the microwave emission comes from a region extending from the chromosphere to the low transition region. We measured the properties of the SSS as well as the degree of circular polarization averaged over the beam of the radio telescope and from this we estimated the magnetic field at the formation level of the radiation.
\end{abstract}
\end{opening}

\section{Introduction}
     \label{intro} 

The microwave range is rich in information on quiet-sun structures in the upper chromosphere, the transition region and the low corona (see \opencite{2011SoPh..273..309S}, for a recent review). The variation of the brightness temperature with frequency provides valuable data for modeling, while the circular polarization gives estimates of the magnetic field.

Due to the limited spatial resolution, useful information can only be obtained with large radio telescopes or interferometers. The chromospheric network was first detected in interferometric data (\opencite{1974SoPh...34..125K}; \opencite{1975MNRAS.173...65K}; \opencite{1978ApJ...224.1043Z}). The first, one-dimensional, imaging observations came from the {\it Radio Astronomical Telescope of the Academy of Sciences} 600 (RATAN-600) in the wavelength range of 2--4 cm, which showed small-scale brightness fluctuations across the quiet-sun (\opencite{1975SvAL....1..205B}; \opencite{1977PAZh....3..550G}; \opencite{1978SoSAO..23...22B}); this phenomenon was called {\it solar radiogranulation}. Thanks to the high spatial resolution (9\arcsec\ by 50\arcsec) at the wavelength of 1.35 cm, it was possible to identify the individual elements of the radiogranulation with bright and dark elements of the chromospheric network visible in the Ca{\sc ii} K line \cite{1982ASSL...96..109G}. In the millimeter range, small-scale structures (SSS) were first observed during the passage of Mercury across the disk of the Sun with the {\it Radio Telescope 22} (RT-22) \cite{1975IzKry..53..121E} and with the {\it Radio Telescope 25} RT-25 by \inlinecite{1975PAZh....1...24K}. \inlinecite{1982ASSL...96..109G} made a comparison with optical images. Typical sizes of the cells obtained by  \inlinecite{1978SoSAO..23...22B} and \inlinecite{1982ASSL...96..109G} were 40\arcsec\ to 50\arcsec\ in the range from 1.96 cm to 10 cm.

Further observations with two-dimensional resolution came from synthesis instruments. \inlinecite{1979ApJ...234.1122K} 
were the first to obtain quiet-sun images at the wavelength of 6 cm with arc-second resolution using the {\it Westerbork Synthesis Radio Telescope} (WSRT). The WSRT quiet-sun images showed a clear association of the quiet-sun microwave emission with the chromospheric network. This conclusion was subsequently verified with the {\it Very Large Array} (VLA) observations at 6 and 20\,cm by \inlinecite{1988ApJ...329..991G} and by \inlinecite{1990ApJ...355..321G} at 3.6\,cm. In the mid 90's the VLA was used for quiet-sun observations in the short cm-range  (1.2, 2.0 and 3.6\,cm) by \inlinecite{1996ApJ...473..539B}, \inlinecite{1997AA...320..993B}, and \inlinecite{1997ApJ...488..499K}. In addition to the chromospheric network, SSS are also associated with other features, such as X-ray bright points, ephemeral active regions, magnetic elements and, possibly, with small coronal loops.

The chromospheric structure is of special interest as an input to the theories of coronal heating (\opencite{2007ApJ...659.1673A}; \opencite{2012A&A...537A.152P}; \opencite{2011A&A...532A.112Z}; see also the review by \opencite{2012SSRv..169..181T}). In recent years significant advances have been made in the study of solar EUV emission with the {\it Hinode} instruments \cite{2012ApJ...750L..25J}, as well as the {\it Atmospheric Imaging Assembly} (AIA)  telescopes aboard the {\it Solar Dynamic Observatory} (SDO) \cite{2012SoPh..275...17L}; nevertheless, these instruments provide no information on the strength of the magnetic field. 

To estimate the magnetic field at the level of the transition region and lower corona, we need to resort to radio observations with the large radio telescope RATAN-600, since the VLA does not have, up to now, sufficient sensitivity to provide accurate measurements of the low circular polarization in the quiet Sun. Despite the lower spatial resolution of RATAN-600, compared to the VLA or to EUV instruments, its sensitivity is high enough to record the SSS in the microwave range. In recent years a new multi-wavelength spectral and polarization equipment has been installed in RATAN-600 (\opencite{2011AstBu..66..190B}; \opencite{2011AstBu..66..205B}), which has a spectral resolution of 1\% and a frequency range from 3\,GHz to 18.2\,GHz. This system uses a high-speed digital data acquisition system with a large dynamic range, thus providing a high signal-to-noise ratio from the level of the quiet Sun up to to that of bursts, in many wavelengths simultaneously.

In this article we present the first results of a study of small-scale quiet-sun structures in the microwave range during the time interval from September 2005 to July 2012, made with the RATAN-600 using its new technical capabilities. The observations and their analysis are described in Section 2; our results are presented in Section 3 and are discussed in the Section 4.  

\section{Observations and Data Reduction} \label{obs}
The observations were made in the southern sector of the antenna system with the periscopic mirror (\opencite{2004R&QE...47..227B}). In this configuration it is possible to make multiple observations at different azimuths. This provides scans along different position angles across the Sun (depending on the declination), together with extended time coverage. In 2012, the polarization characteristics of the receiving equipment were improved. The former spiral right and left circular polarization feeds, which were shifted from the focus, were replaced by wide-range sinus-type feeds which have a common phase center for both the right and left circular polarizations \cite{2011AstBu..66..190B}. This eliminated wavelength-dependent aberrations due to the shift from the focus.

\begin{figure}[h]
\begin{center}
\includegraphics[width=0.8\textwidth]{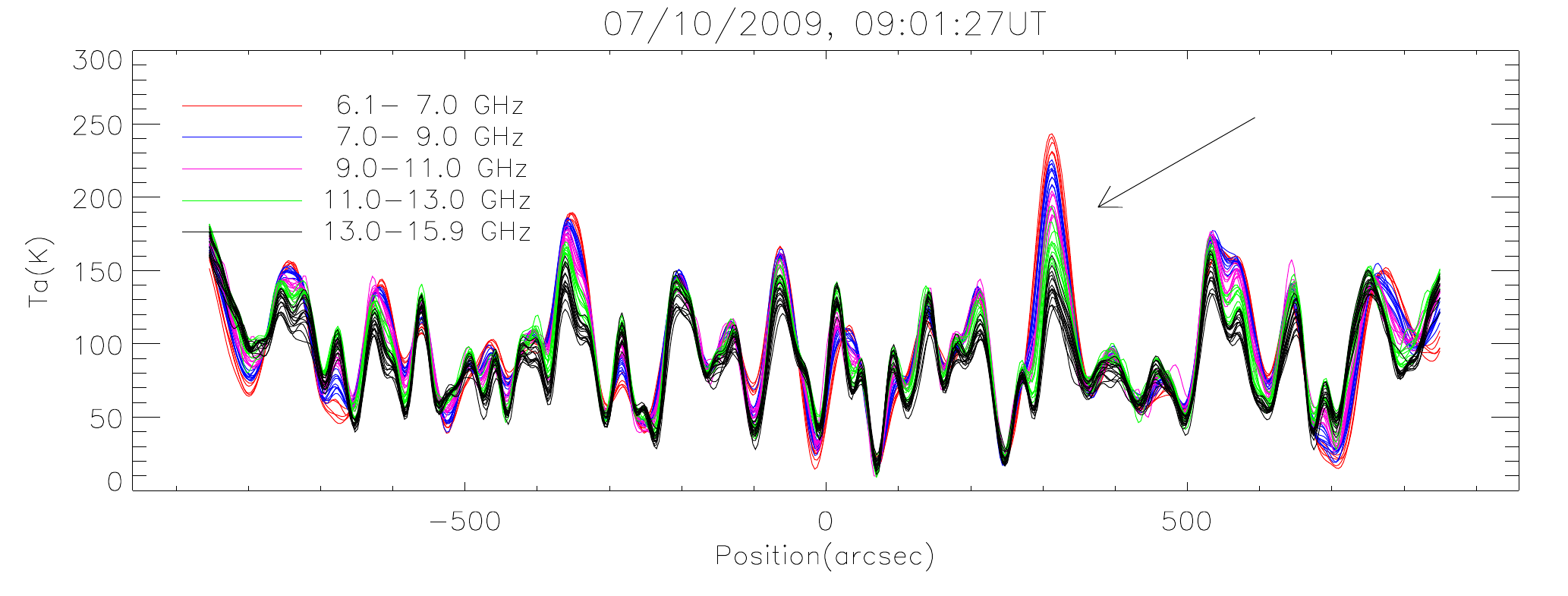}\\
\includegraphics[width=0.26667\textwidth]{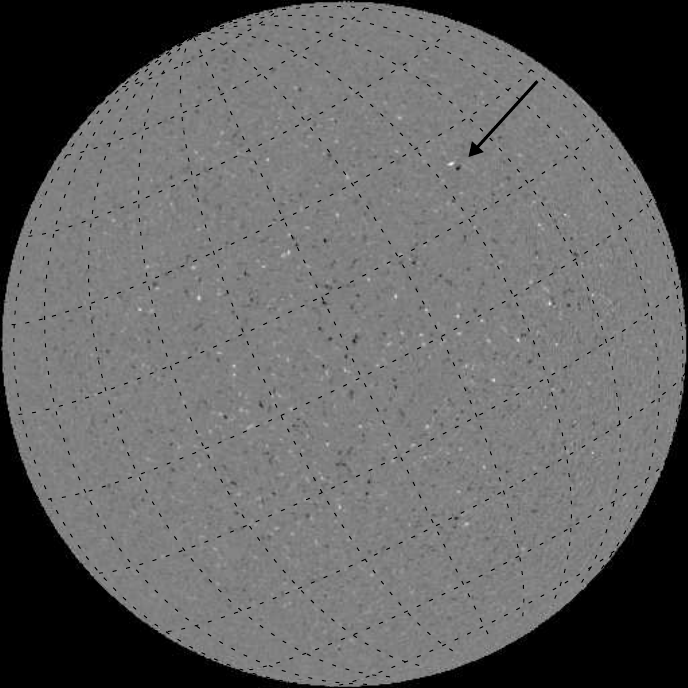}\includegraphics[width=0.26667\textwidth]{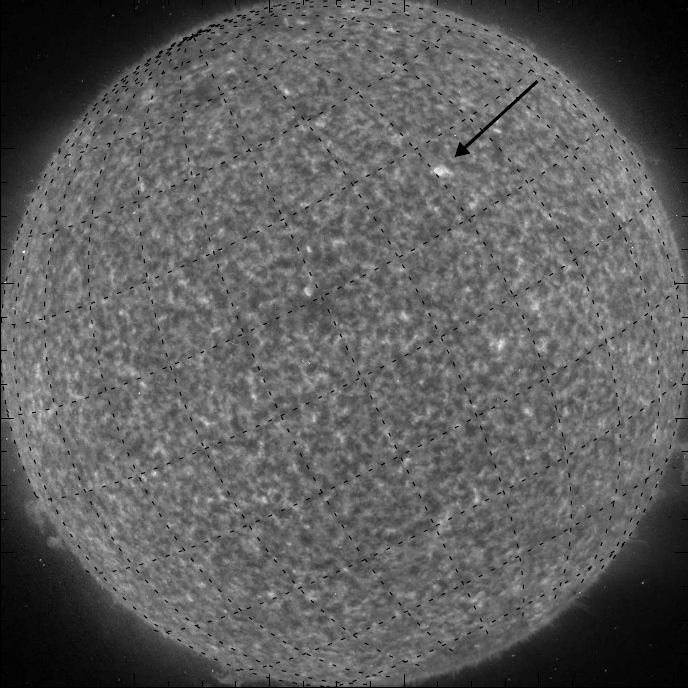}\includegraphics[width=0.26667\textwidth]{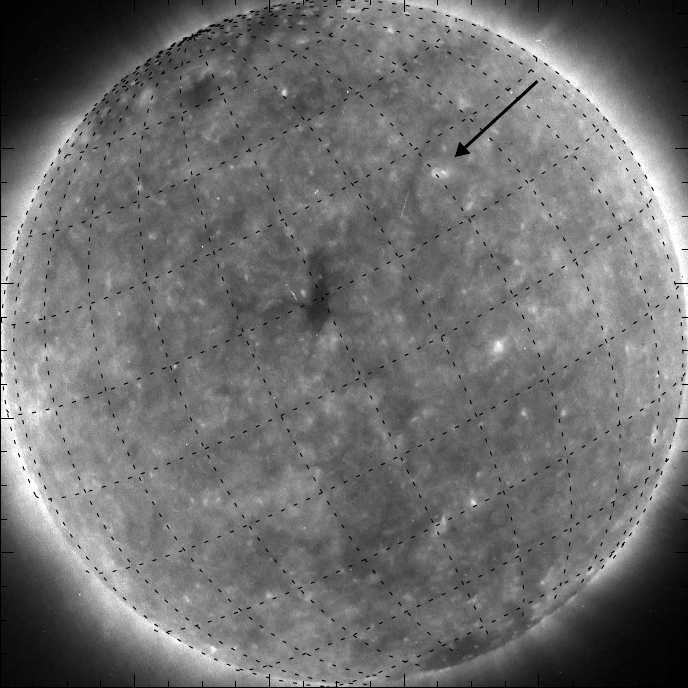}
\end{center}
\caption{Example of multi-wavelength 1D images (scans) of the quiet-sun obtained with the RATAN-600 radio telescope on 7 October 2009 (top). In the bottom row, from left to right, images from the {\it Solar and Heliospheric Observatoty} (SOHO) {\it Michelson Doppler Imager} (MDI), and the {\it Extreme ultraviolet Imaging Telescope} (EIT) at 304\,\AA\ and 171 \AA\ are shown, oriented in the celestial EW-NS direction. The arrow marks the brightest source on the radio scans and the images.}
\label{fig01}
\end{figure}

The radio telescope has a large effective area (300--600\,m$^2$) in the centimeter range. It provides a flux sensitivity of $\sim0.01$\,sfu, with a position accuracy of about 1\arcsec. It covers a broad frequency band with high spectral resolution, it has a relatively wide range of coverage in time (07:00\,UT to 11:00\,UT), and high-precision measurement of the polarized signal, 0.05 to 0.2\%. 

\begin{figure}
\begin{center}
\includegraphics[width=.5\textwidth]{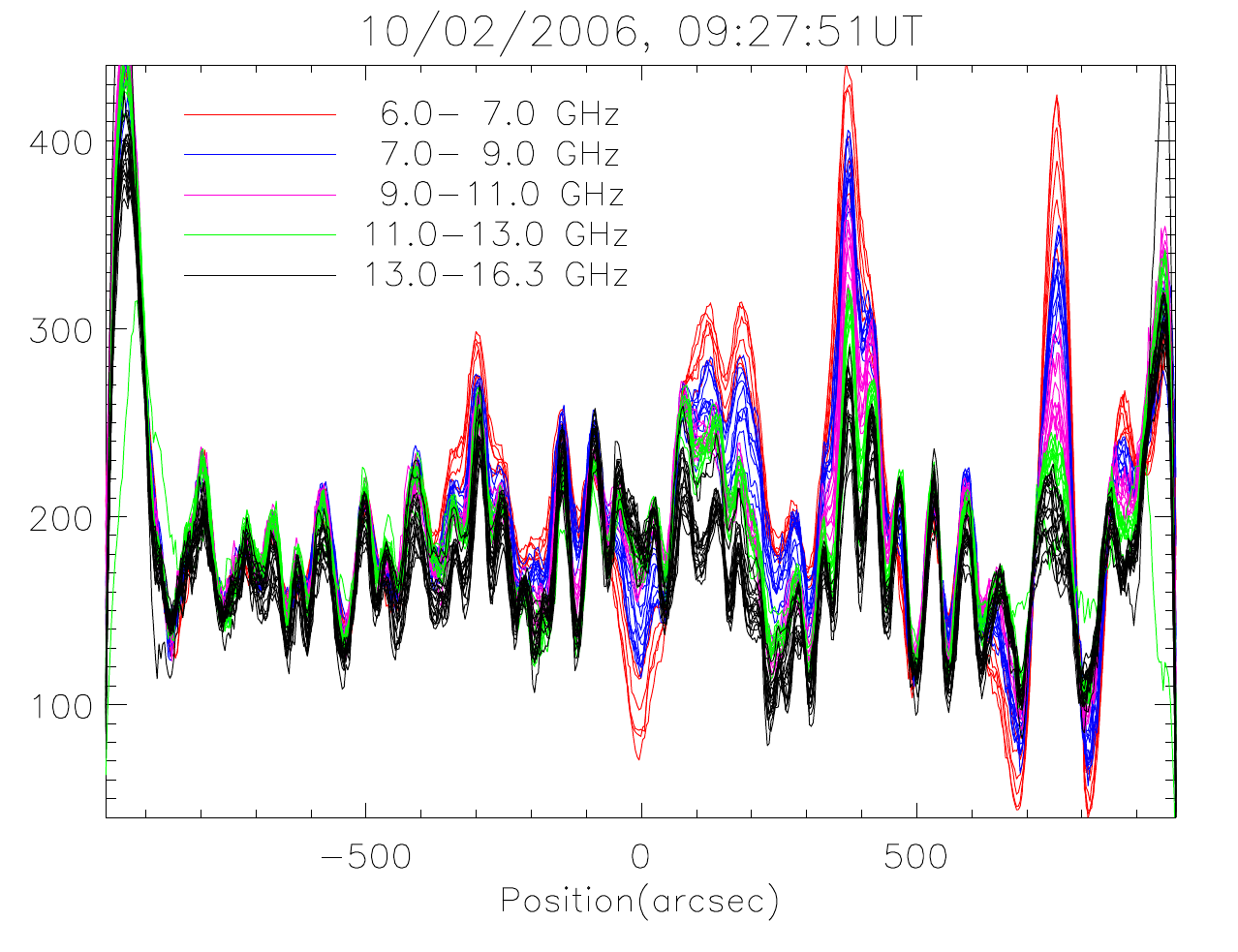}\includegraphics[width=.5\textwidth]{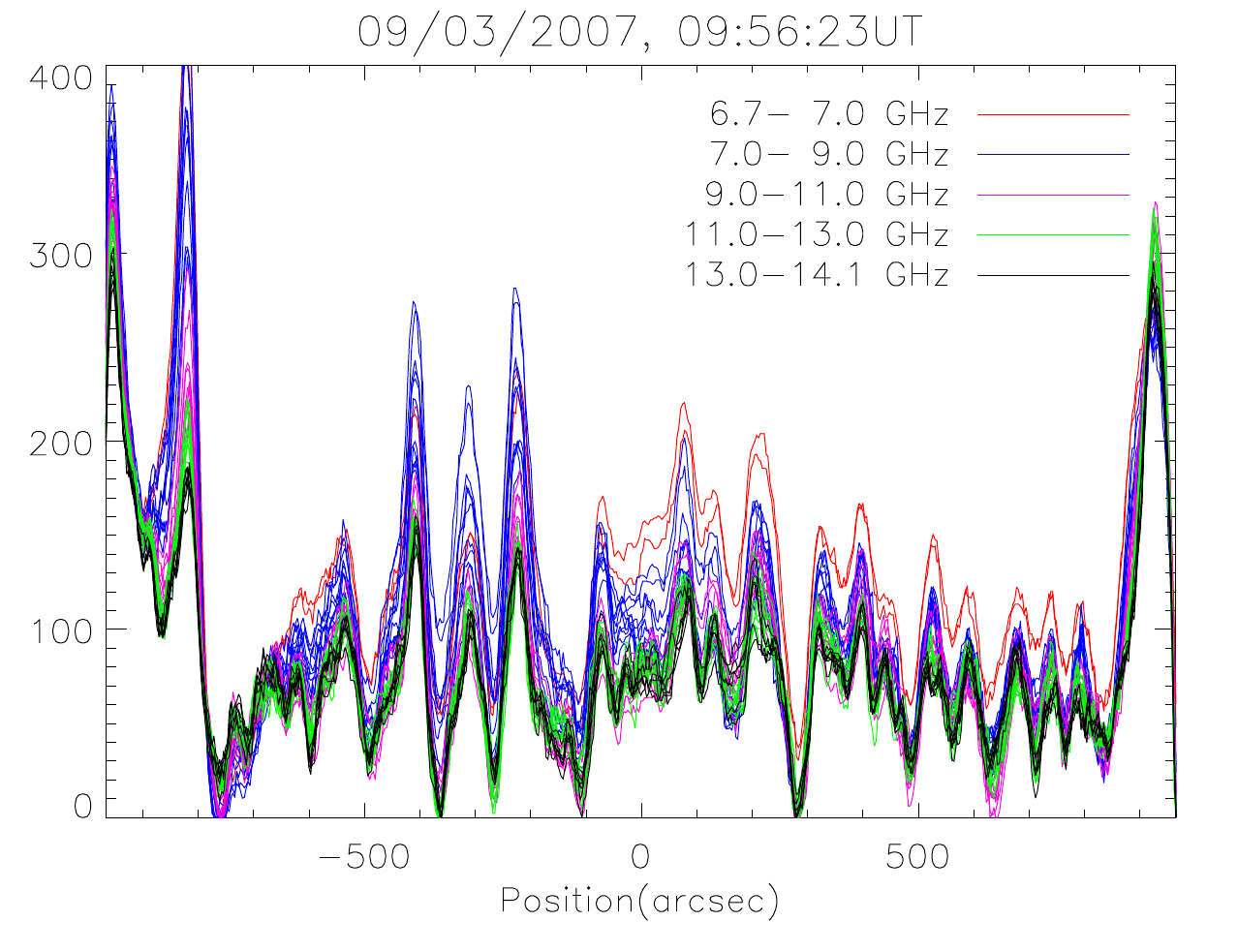}\\
\includegraphics[width=.5\textwidth]{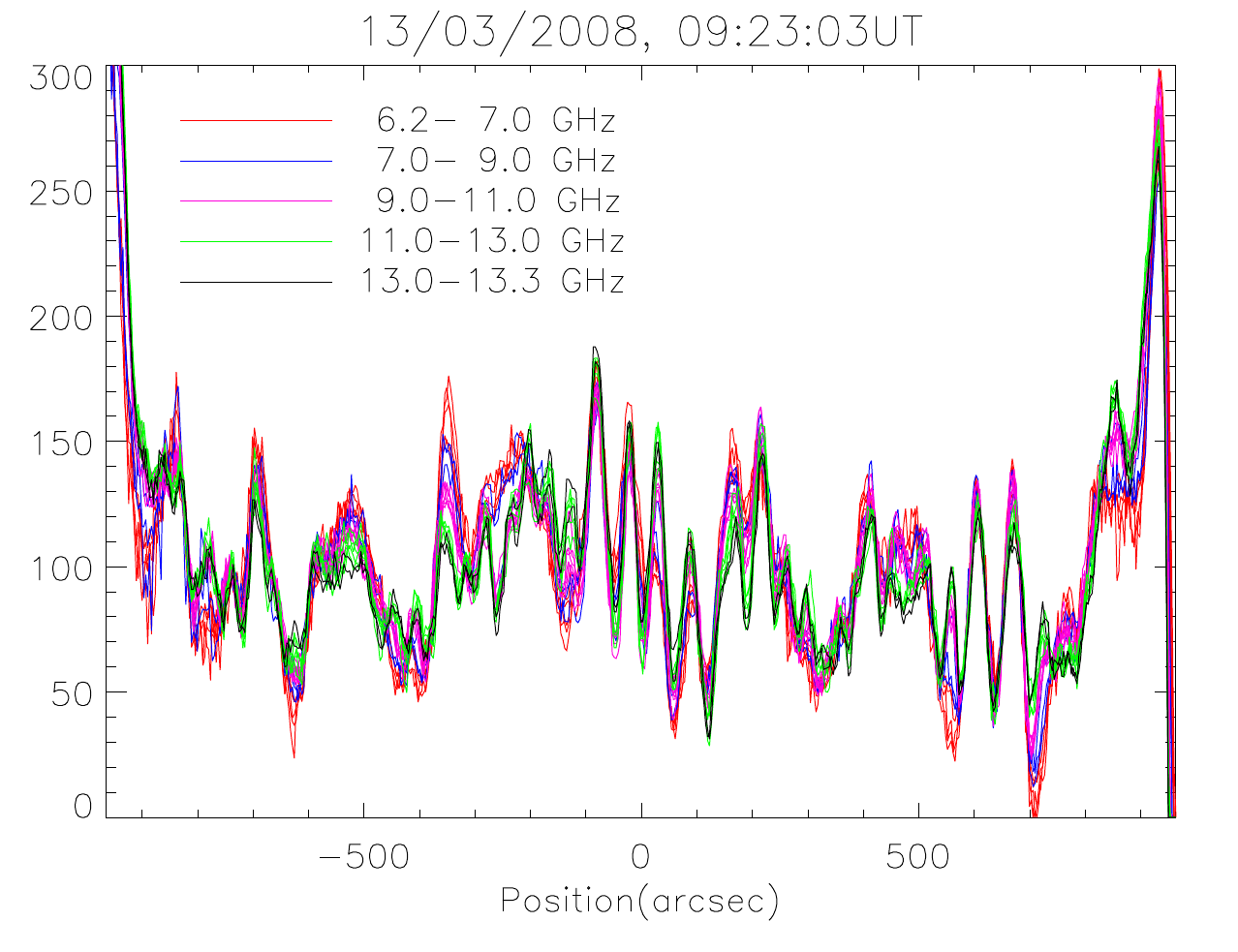}\includegraphics[width=.5\textwidth]{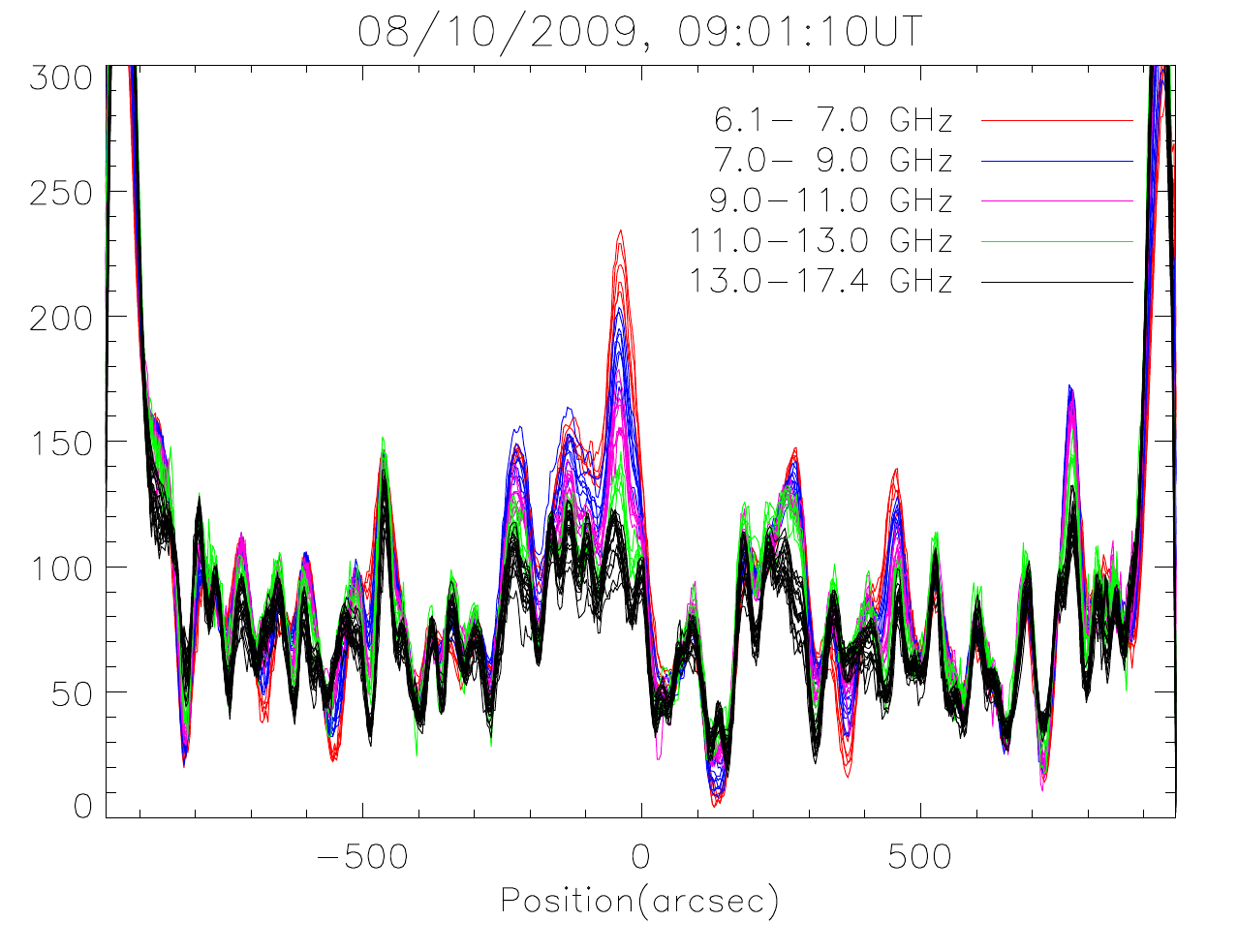}\\
\end{center}
\caption{Examples of recordings of solar SSS in the range of 6 to 18 GHz during the period of solar activity minimum, 2005 to 2009. The intensity scans are shown after subtraction of the large-scale background component. Scans in several wavelengths in the range of 6--16.4 GHz are presented. The variations are 0.01 to 0.03 of the quiet-sun antenna temperature.}
\label{fig02}
\end{figure}

In the early observations of quiet-sun small-scale structures obtained with RATAN-600 (\opencite{1975SvAL....1..205B}; \opencite{1978SoSAO..23...22B}), the antenna system of the northern sector was used; in that system the vertical size of the beam was only 6 times larger than the horizontal one. Since 2005, regular solar observations with RATAN-600 are made, using the southern sector with a periscopic mirror. The beam pattern has a knife-edge shape (for 
example, at 2 cm the beam size is  17\arcsec\ by 15\arcmin).  Assuming a uniform distribution of the network cells, one could suggest that a strong smoothing effect should exist due to the inclusion of at least 12--15 cells in the vertical beam pattern. However, as can be seen from Figures \ref{fig01} and \ref{fig02}, the SSS is well depicted in the one-dimensional (1D) scans over a broad frequency range.

The digital multi-channel data recording system \cite{2011AstBu..66..371B} provides an increased dynamic range, which extends from the instrumental noise level (antenna temperature of about 300 to 500\,K) up to the high-power signals 
\begin{figure}[h]
\begin{center}
\includegraphics[width=\textwidth]{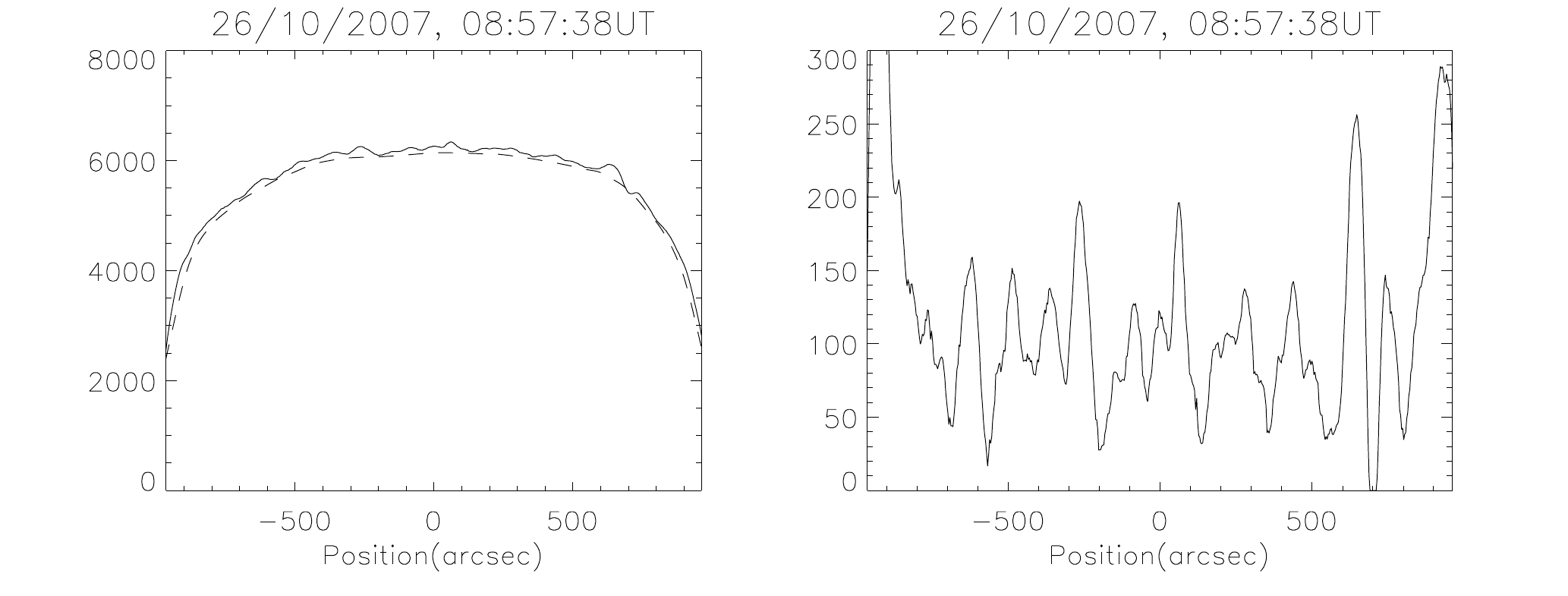}\\
\end{center}
\caption{Large-scale component subtraction to enhance small-scale structures. The left panel shows the original scan and the background component (dashed line), while the right panel shows the residual structure after the subtraction}
\label{fig03}
\end{figure}
during flares (about 200 to 500$\times10^3$\,K). This corresponds to about 30 to 80 $\times T_a$, where $T_a$ is the level of the quiet-sun antenna temperatures.
To separate the quiet-sun structures the following method is used: 
\begin{enumerate}
\item Calibration of the multi-wavelength data. 
\item Subtraction of the large-scale background component, including the quiet-sun emission (Figure \ref{fig03}); this leaves some residual background emission near the limbs. 
\item High-pass filtering to reduce small-scale noise with a characteristic angular size much smaller than the resolution.
\end{enumerate}

\section{Results}

\subsection{Temporal Characteristics}
\begin{figure}[h]
\begin{center}
\includegraphics[width=0.8\textwidth]{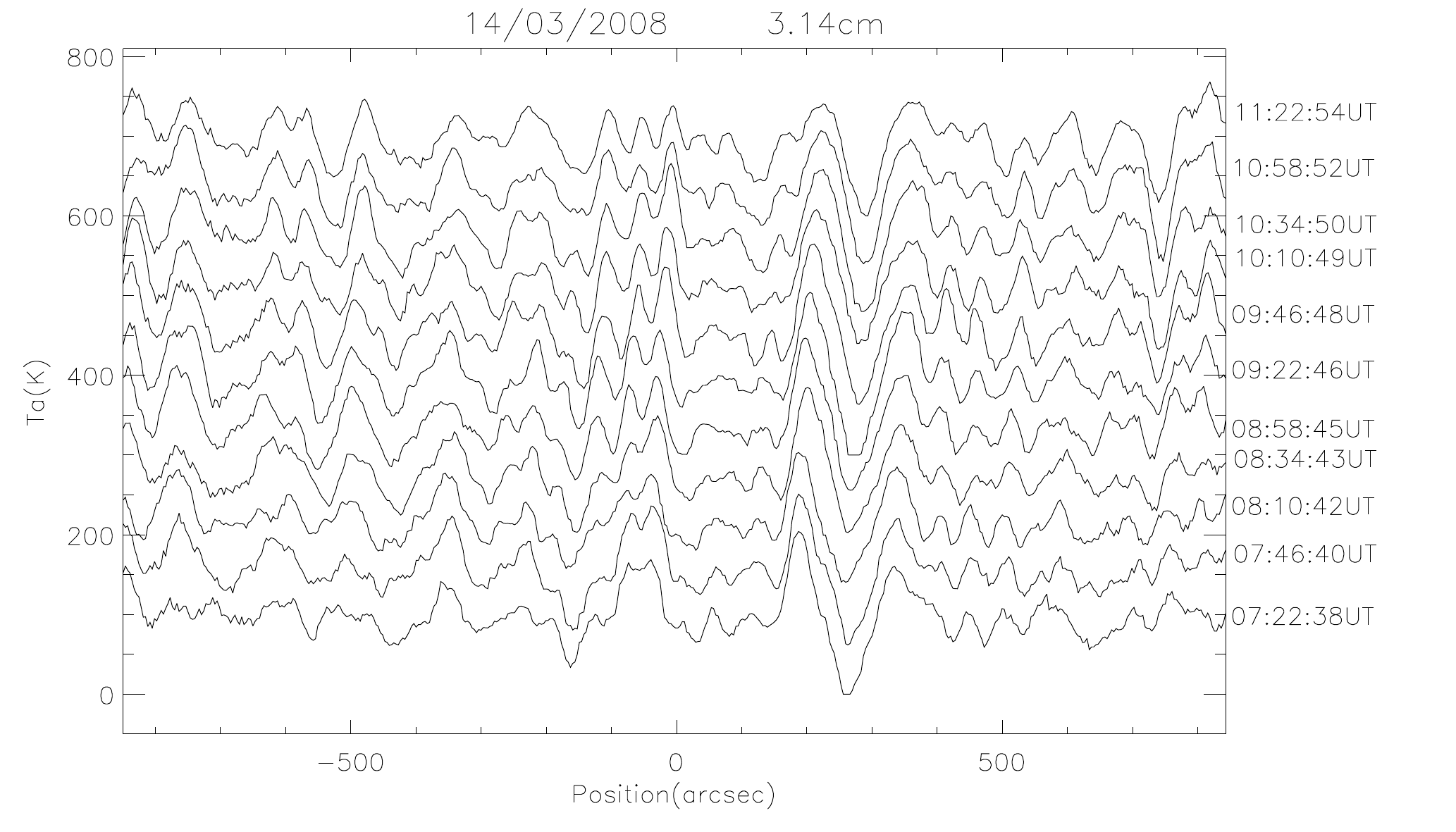}\\
\end{center}
\caption{Changes in the SSS during 4 hours, from observation at different azimuths. The time interval between consecutive scans is 24 minutes. The gradual drift of the sources to the right is due to the solar rotation}
\label{fig05}
\end{figure}

Observations at different azimuths allowed us to study the stability in time of the SSS. Figure \ref{fig05} shows four-hour long observations in 11 azimuths obtained on 14 March 2008, with a cadence of about 24 minutes. Note that these observations were taken near the time of the vernal equinox, so that the orientation of the scans with respect to
the solar E--W direction is practically the same. Taking into account the rotation of the Sun, a high degree of correlation ($> 82$\%) is found between consecutive scans, and up to 60\% between observations separated by a few hours. Thus the SSS are a long-lived solar phenomenon, rotating with the Sun. The average brightness of the SSS is about 1--1.5\% of that of the quiet-sun

\subsection{Comparison of RATAN-600 Data with SDO/AIA Data}
\begin{figure}[h]
\begin{center}
\includegraphics[width=0.8\textwidth]{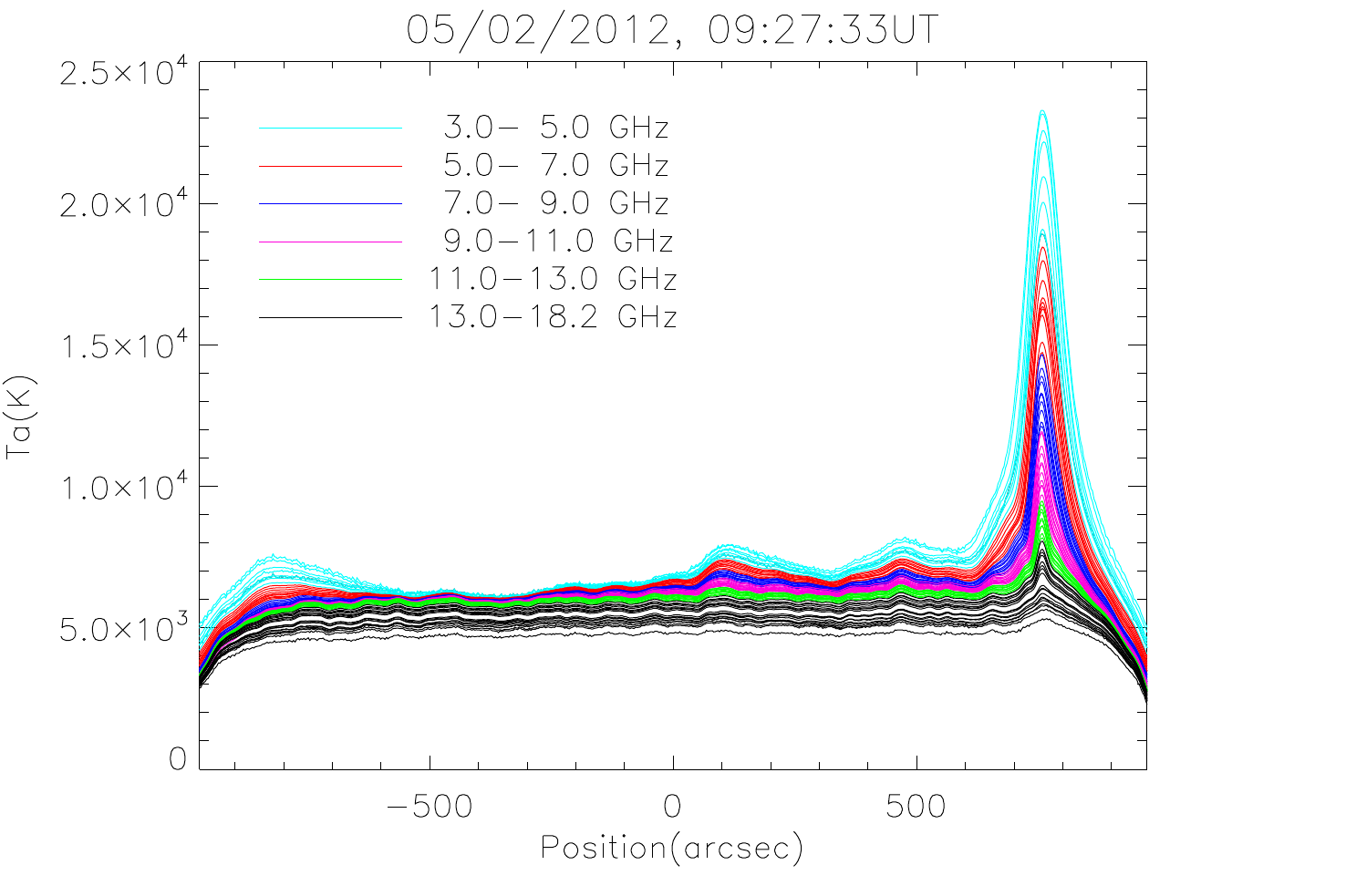}\\
\end{center}
\caption{Multi-wavelength RATAN-600 1D solar scans for5  February 2012}
\label{fig06}
\end{figure}

In order to localize the radio structures on the disk, we compared the RATAN-600 1D scans with optical and EUV images. After the new polarization feeds were installed (see Section \ref{obs}), solar activity was close to its maximum therefore, there were not many days devoid of active regions. For our analysis we selected the date of 5 February 2012, when the SSS is well visible despite the fact that an active region was present near the west limb (Figure \ref{fig06}).

\begin{figure}
\begin{center}
\includegraphics[width=0.9\textwidth]{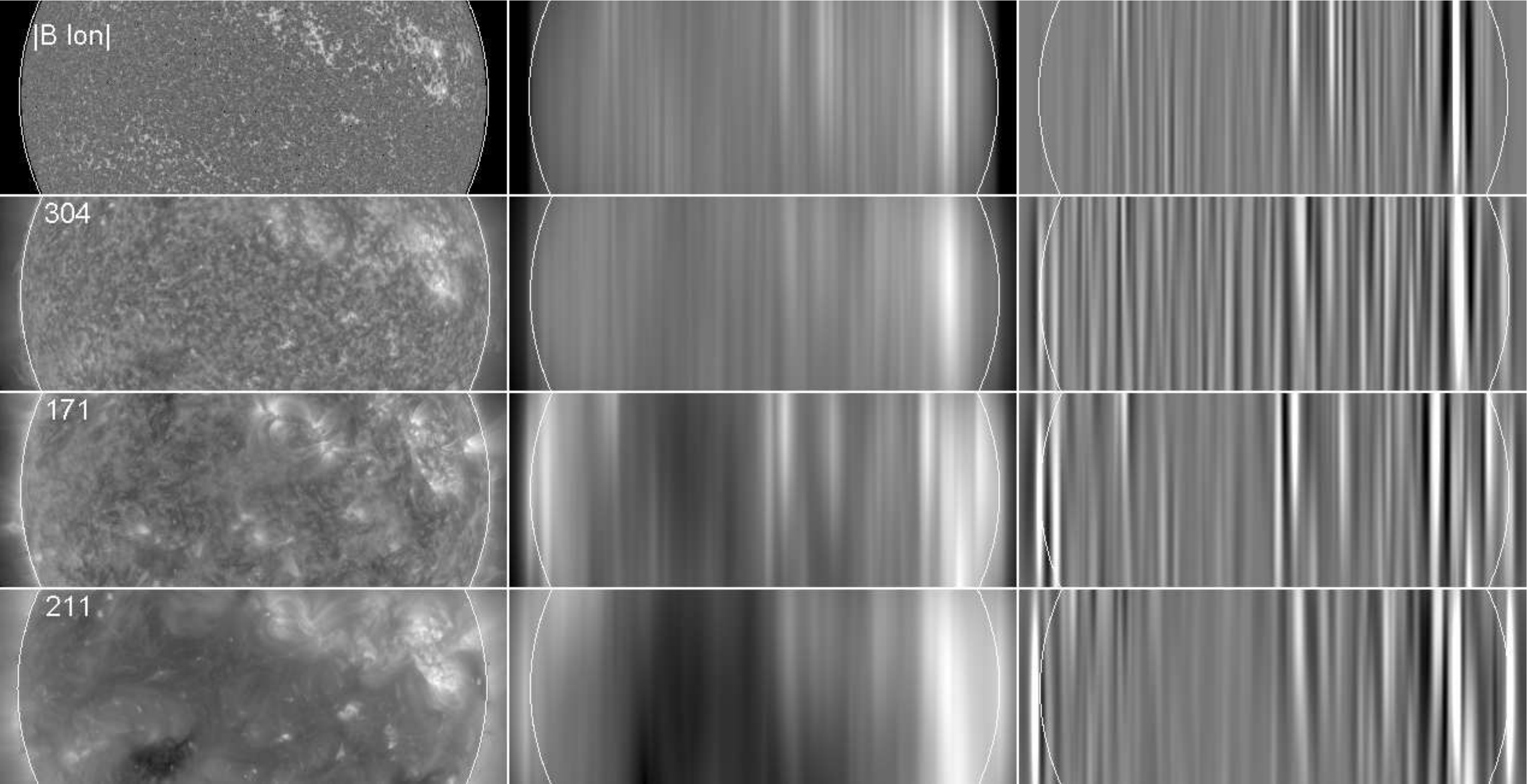}\\
\end{center}
\caption{Left column: Original SDO images for 5 Februry 2012 (from top to bottom: absolute value of the  longitudinal magnetic field fromthe {\it Helioseismic and Magnetic Imager} (HMI), images in 304, 171 and 211\,\AA\ bands from AIA). Central column: similar to the left column, after convolution with the RATAN-600 beam pattern at 1.65\,cm. Right column: the result of high-pass filtering in the E--W direction. The images are centered at the disk center and are oriented in the celestial E--W and N--S direction. The size of the region shown is 2100\arcsec\ by 800\arcsec. White arcs mark the photospheric limb}
\label{fig07}
\end{figure}

SDO provides the best set of images for comparison with the RATAN-600 data. They have a high cadence which permits to select  images very close to the RATAN-600 observations and cover a large range of heights, from the chromosphere to the low corona. The first step in the comparison was the convolution of the SDO images with the RATAN-600 beam. Subsequently, we removed large-scale structure in the E--W direction by high-pass filtering with a gaussian of 60\arcsec\ width. Figure \ref{fig07} shows in the left column the original SDO images in selected SDO bands, the middle column shows the same images convolved with the instrumental beam at the shortest wavelength (1.65\,cm), while the result of the high-pass filtering is shown in the right column. Note that small-scale structures persist, in spite of the smoothing with the RATAN-600 beam. 

\begin{figure}
\begin{center}
\includegraphics[width=0.9\textwidth]{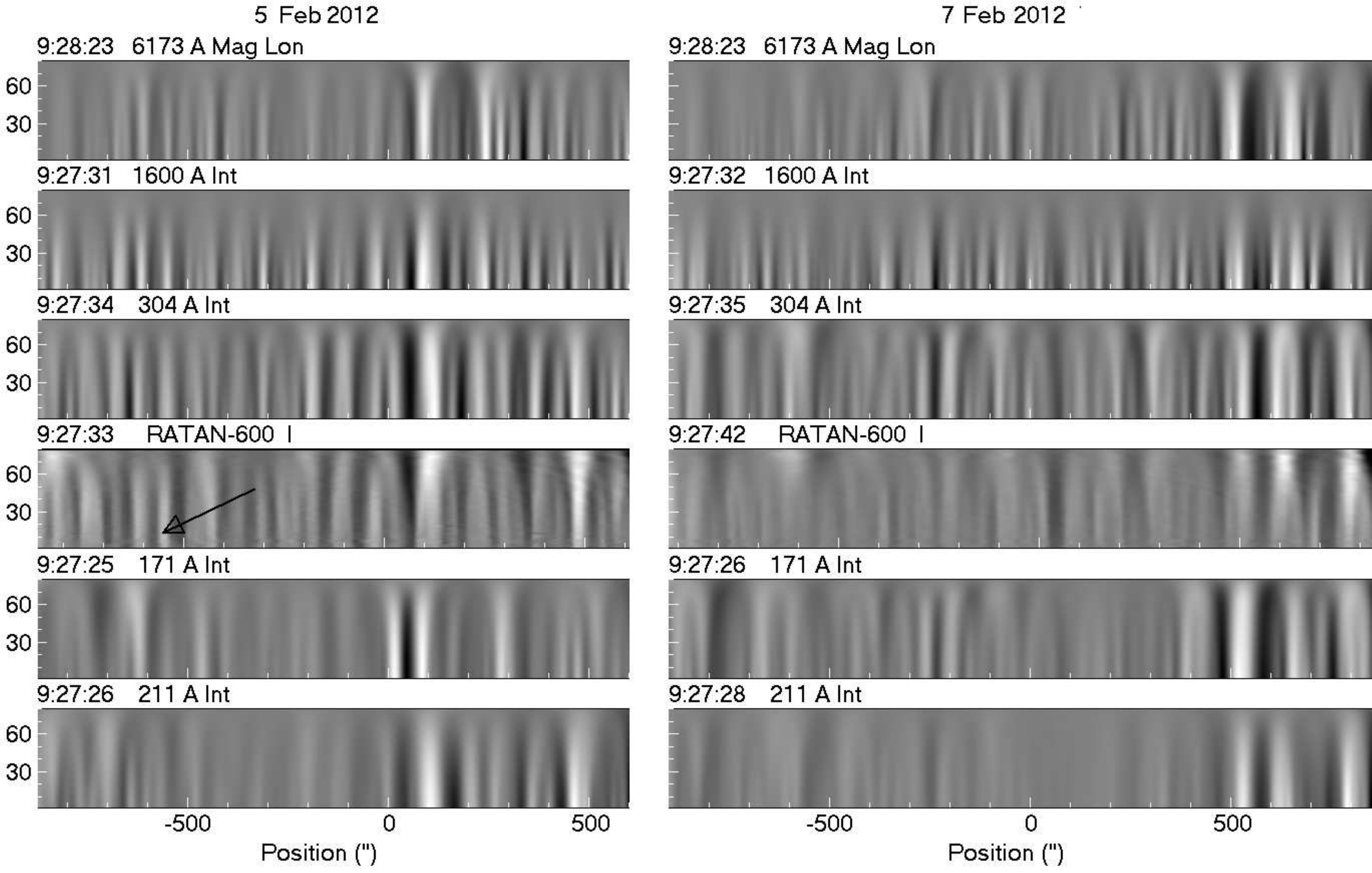}\\
\end{center}
\caption{Comparison of images of the RATAN-600 antenna temperature as a function of position and frequency with simulated SDO images ($|B_\ell|$, intensity in the 1600, 304, 171 and 211\,\AA\ bands), for 5 February (left) and 7 February 2012 (right). The range of positions was selected such as to avoid the active region near the West limb. The ordinate is the frequency channel, which is almost linear in frequency with channel 1 at 18.2\,GHz (1.65\,cm) and channel 80 at 3\,GHz (10\,cm). The arrow in the RATAN-600 data of 5 February points to the feature analyzed in Section \ref{feature}}
\label{sdorat}
\end{figure}

The change of the RATAN-600 resolution, as well as possible changes of the radio structures as a function of wavelength due to the change of the opacity and the subsequent change of formation height, complicates the comparison of the two data sets. To avoid the problem with the change of resolution, we simulated the effect of the RATAN-600 beam on the SDO images. In the images shown in Figure \ref{sdorat}, each row is the convolution of an SDO image by the radio beam at a particular frequency; thus, they simulate what SDO would observe if it operated in the same way as RATAN-600.  These images are directly comparable to images of the RATAN-600 antenna temperature as a function of position and frequency, also shown in Figure \ref{sdorat}. The RATAN-600 data were also subjected to high-pass filtering for consistency with the SDO data. In addition to the 5 February 2012 data we also present data for 7 February.

The similarity of the RATAN-600 data to the SDO data, particularly in the 1600 and 304\,\AA\ bands is striking; there is practically a peak-to-peak correspondence in the entire frequency range for both days. In order to obtain a more quantitative result, we computed the cross-correlation coefficient between the antenna temperature at each radio frequency and the corresponding SDO simulated scan. The average for all frequencies is given in Table \ref{cross}. 

\begin{table}[h]
\caption{Cross-correlation between RATAN-600 and SDO data}
\label{cross}
\begin{tabular}{rcc}
\hline 
 Band & 5 February & 7 February\\
$|B_\ell|$ & 0.34 & 0.23 \\ \\
1600 & 0.58 & 0.64 \\
  304 & 0.71 & 0.74 \\
  171 & 0.25 & 0.52 \\
  193 & 0.31 & 0.63 \\
  211 & 0.43 & 0.64 \\
  131 & 0.18 & 0.61 \\
  335 & 0.46 & 0.68 \\
   94 & 0.36 & 0.66 \\
\hline 
\end{tabular}
\end{table}

The results show that the highest correlation of the radio data is with the 304\,\AA\ band (0.71 and 0.74 for February 5 and 7 respectively), formed in the low transition region at a temperature of about 10$^5$\,K. The next high correlation is with the 1600\,\AA\ band (0.58 and 0.64), formed in the low chromosphere. Relatively low is the correlation with the absolute value of the longitudinal magnetic field, $|B_\ell|$, probably due to the fact that the radio emission is related to the absolute value of the magnetic field rather than to its longitudinal component. We note, for comparison, that the correlation between the 304\,\AA\ band and the $|B_\ell|$ images is not high either; its  value is 0.44 and 0.34 for the two days considered. The correlation drops significantly as we go to the 171\,\AA\ band, formed in the middle to upper transition region (at $\sim 10^6$\,K). There is a slight increase of the correlation with the SDO bands formed in the low corona, but this is probably due to the plage rather than the quiet-sun component of the emission; indeed, as can be seen in Figure \ref{sdorat}, most microwave features in the eastern hemisphere, which is devoid of plage regions, are not visible in the 211\,\AA\ band. 

\begin{figure}
\begin{center}
\includegraphics[width=\textwidth]{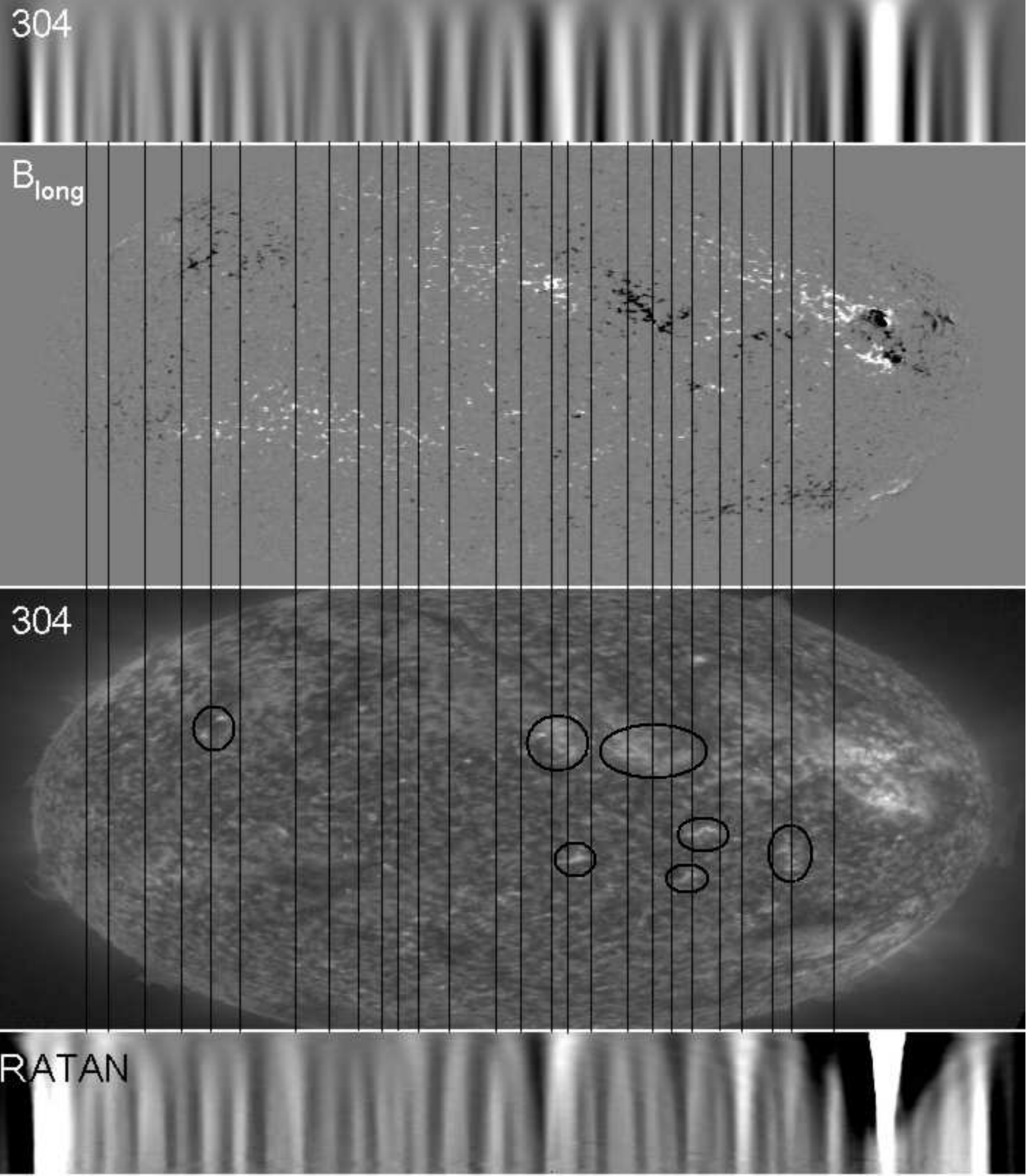}\\
\end{center}
\caption{Comparison of the RATAN-600 position-frequency data (bottom) of 5 February 2012 with the AIA image in the 304\,\AA\ AIA band and the HMI image of $B_\ell$, both compressed in the N--S direction. For reference, the top panel shows the 304\,\AA\ data in a form simulating the RATAN data. Vertical lines mark the 1D position of the SSS peaks.}
\label{compare}
\end{figure}

The identification of the peaks seen in the 1D scans with solar features presents inherent difficulties, due to the confusion introduced by the integration in the N--S direction. Such a comparison is shown in Figure \ref{compare}, where the RATAN-600 positional-spectral image and the corresponding AIA 304\,\AA\ simulated image ({\it cf.} Figure \ref{sdorat}) are shown, together with the original 304\,\AA\ and $B_\ell$ images. Apart from the active region near the limb, the comparison shows a number of cases, marked by ellipses in the 304\,\AA\ image, where the SSS peaks are associated with spotless plage or small bipolar regions. In most cases, however, there is no clear association of the peaks with distinct structures either in the 304\,\AA\	 image or in $B_\ell$. Apparently, such peaks are the sum of several individual structures.

\subsection{Spectral Properties of the SSS}
\label{feature}
\begin{figure}
\begin{center}
\includegraphics[width=.8\textwidth]{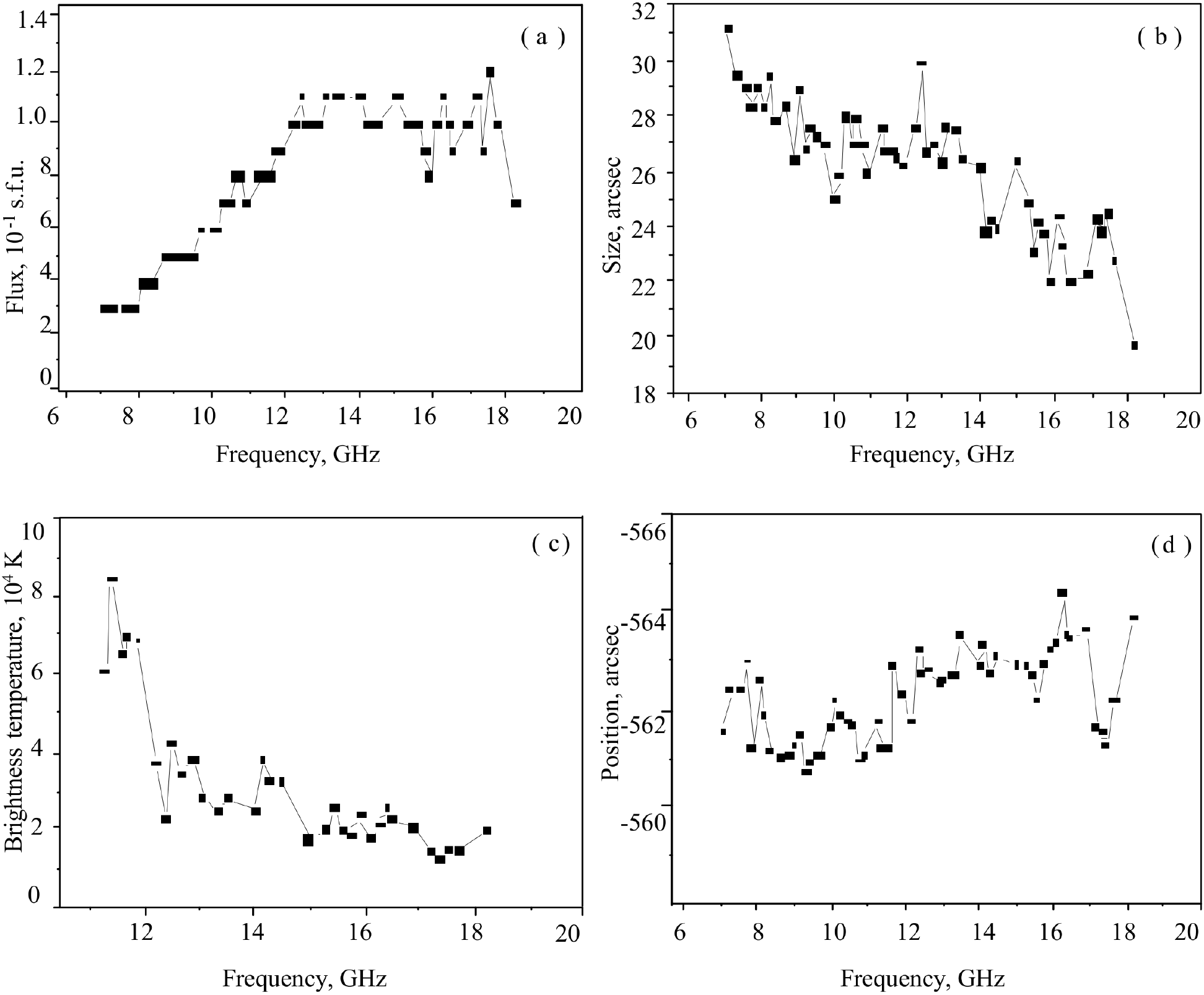}\\
\end{center}
\caption{The flux (a), the size (b), the brightness temperature (c) and the position (d) of the feature marked by the arrow in Figure \ref{sdorat} as a function of the frequency.}
\label{fig11}
\end{figure}

We investigated the spectral properties of an individual peak, marked with the arrow in Figure \ref{sdorat}, using a least square fit of the antenna temperature profile with a Gaussian model \cite{GAR1997}. We thus measured the flux, the size, the brightness temperature and the position of the feature as a function of frequency (Figure \ref{fig11}). Note that the brightness temperature could not be reliably computed below 11 GHz.

The flux density of this feature demonstrates a clear steady increase with frequency, which stabilizes around 12\,GHz (2.5\,cm). The size decreases from about 30\arcsec\ at low frequencies to about 20\arcsec\ at high frequencies; this variation should be due to the combined effect of the change of instrumental  resolution and real structural changes, as mentioned previously. The brightness temperature shows little change at high frequencies (from 18 to 13\,GHz, 1.7 to 2.3\,cm), with a value of about 20$\times 10^3$\,K, subsequently it increases to about 70$\times10^3$\,K at 11\,GHz; this increase could be due to the shift of the level of formation of the radiation into the low transition region. The position of the peak is constant in most of the frequency range, but below 6 GHz it is shifting eastwards (see also Figure \ref{sdorat}). This could be due to the loss of resolving power at low frequencies and subsequent merging with other peaks.

\begin{figure}
\begin{center}
\includegraphics[width=\textwidth]{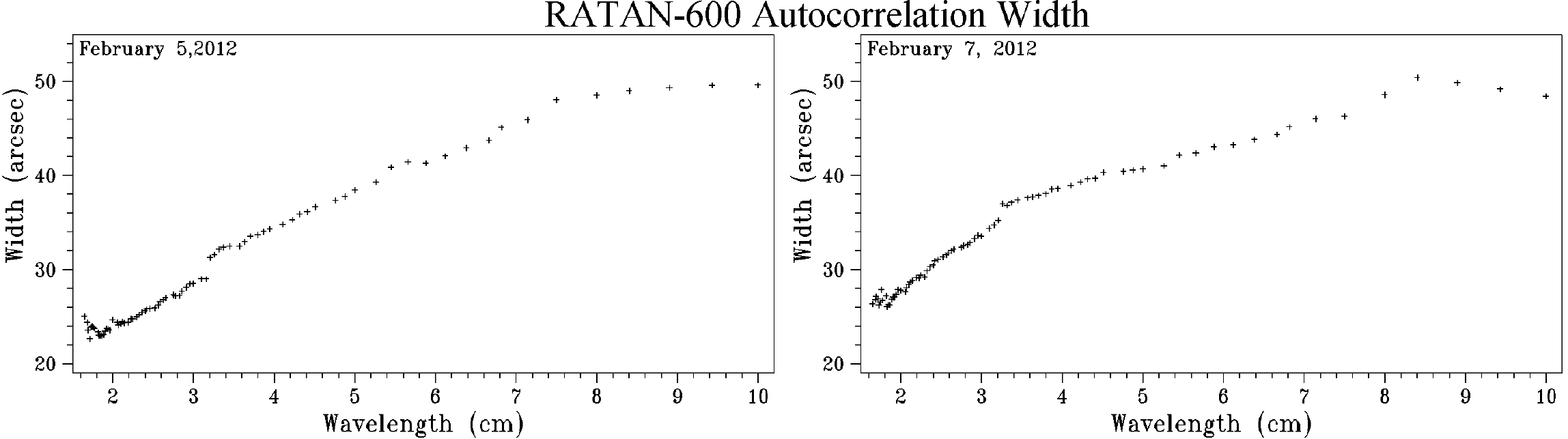}\\
\end{center}
\caption{The width of the autocorrelation function as a function of frequency.}
\label{autowid}
\end{figure}

A more global estimate of the characteristic size is provided by the width of the autocorrelation function in the spatial domain. Our computations (Figure \ref{autowid}) give an autocorrelation width of 25 to 30\arcsec\ at 1.65\,cm, increasing at longer wavelengths to $\sim 50$\arcsec; these values are consistent with those quoted in the previous paragraph. At shorter wavelengths the increase of the width is almost linear, as expected for a structure small compared to the beam size; however, above $\sim7.5$\,cm the value of the width stabilizes. This is also seen in Figure \ref{sdorat} and it probably reflects a real change in structure, apparently due to the rise of the level of formation of the radiation.

\subsection{Polarization and Estimate of the Magnetic Field of Individual Features}
\begin{figure}[h]
\begin{center}
\includegraphics[width=0.8\textwidth]{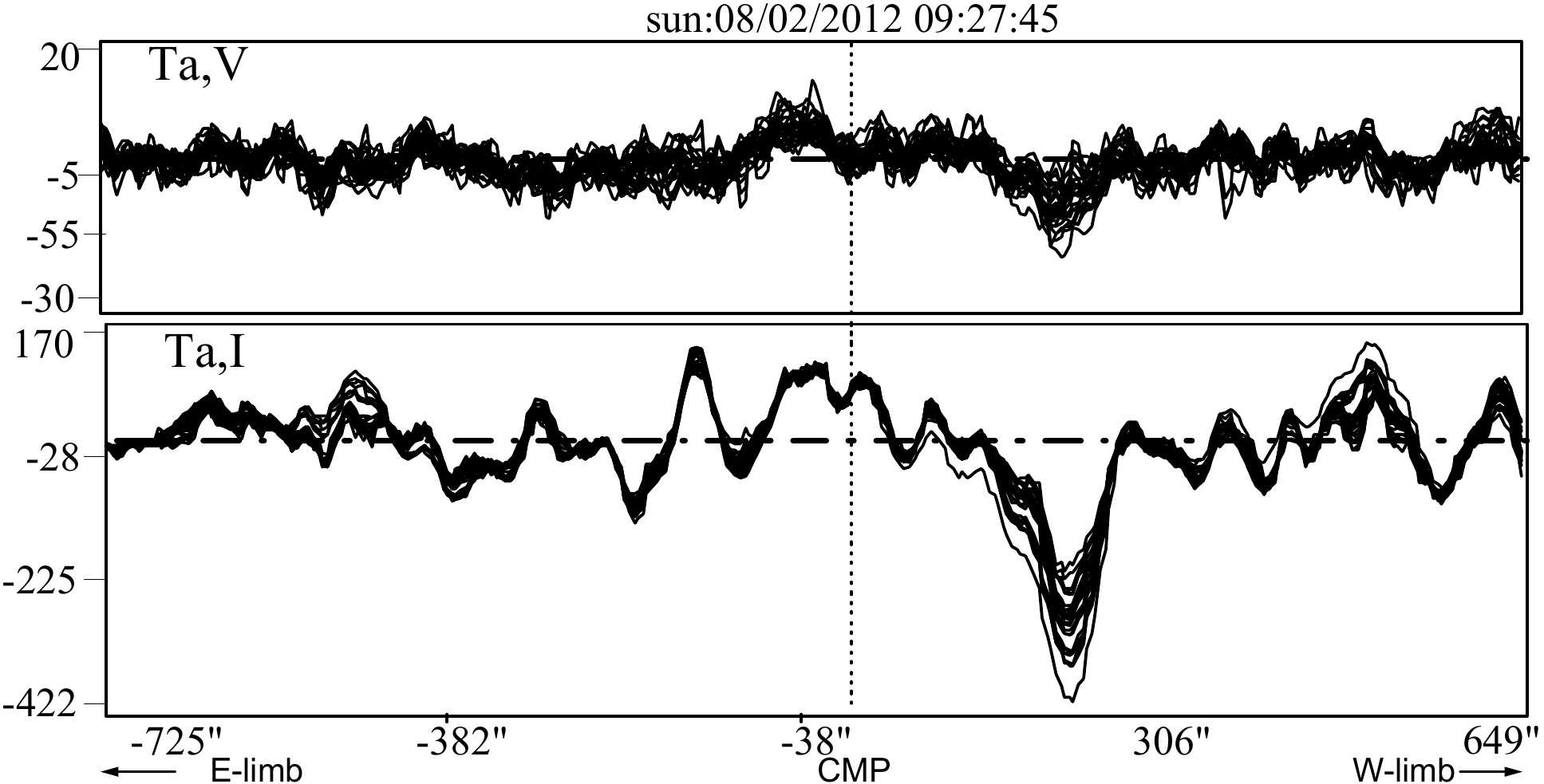}\\
\end{center}
\caption{Stokes V (top) and I (bottom) data for 8 February 2012 in the wavelength range of 1.82 to 3.26\,cm at 27 frequencies. The antenna temperature scale of the V plot is expanded by a factor of 10.}
\label{iv}
\end{figure}

It is interesting to obtain estimates of the magnetic field at altitudes of the lower corona, using polarization observations of the SSS. In Figure \ref{iv} multi-wavelength observations of the SSS in Stokes parameters I and V are shown for 8  February 2012, processed as described above. We note that the polarized radiation is above the background noise, higher than the $\pm 3 \sigma$ level. The estimated maximum value of the degree of polarization is about 4--5\%

\begin{figure}
\begin{center}
\includegraphics[width=0.6\textwidth]{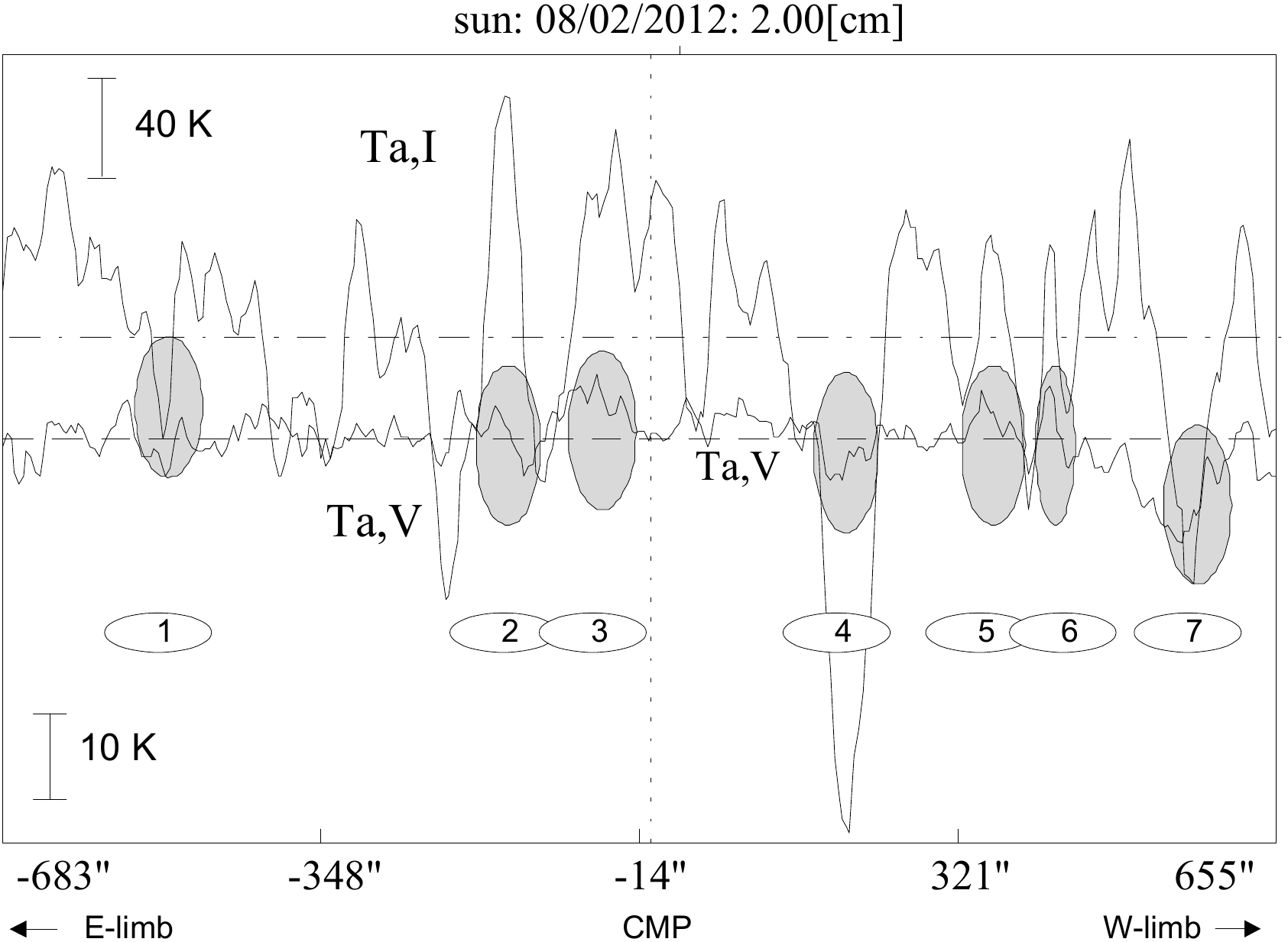}\\
\end{center}
\caption{Analysis of several individual structures in the 2\,cm scan. The circles mask the sources for which the magnetic field was computed}
\label{ivplots}
\end{figure}

To analyze the properties of SSS polarized emission we used the observations at the of 2\,cm of 8 February 8 2012  (Figure \ref{ivplots}). Seven radio structures were selected, with values of their polarized signal exceeding the 3$\sigma$ confidence level. Table \ref{Table01} gives the values of the polarization degree and the magnetic field, computed under the assumption of thermal free-free emission and homogeneous magnetic field; in this case the magnetic field is,
\[B=5400 p /\lambda\]
where $p=V/\Delta I$ is the degree of polarization, $\Delta I$ is the intensity difference with respect to the background and $\lambda$ the wavelength in cm. The obtained values of the magnetic field, which are in the range of 40 to 200 Gauss, should be considered as a weighted average, due to integration within the instrumental beam.

\begin{table}
\caption{Values of the polarization degree and the computed magnetic field for the structures marked in Figure \ref{ivplots}}\label{Table01}
\begin{tabular}{lccccccc}
\hline 
N$^o$  &	1&	2&	3&	4&	5&  6&  7\\
\hline 
P (\%) &3.75&2.5&5.8&1.4&6.25&8.5&7.1\\
B (G)  &100  &67&155& 37&170 &230&190\\
\hline
\end{tabular}
\end{table}

\section{Summary and Conclusions}
In spite of the large width of the RATAN-600 beam in the N--S direction, small scale structures are detected in the 1D scans, in the wavelength range of 1.65 to 10\,cm. The sensitivity of the instrument is sufficient to detect flux variations of $\sim$ 0.01 sfu. The SSS, with a brightness of 1 to 3\% of the quiet-sun level, are stable over several hours, with a characteristic time scale of about one day. Their observed characteristic size increases with wavelength from about 25\arcsec\ to about 50\arcsec, as the resolution decreases; after $\sim7.5$\,cm the characteristic size stabilizes, which could be an indication of a real structural change.

A comparison of the microwave SSS with SDO images, treated in a way that simulates the behavior of RATAN-600, showed an almost one-to-one correspondence between the microwave structures and those seen in the 304\,\AA\ AIA band, with a somewhat inferior correlation with the 1600\, \AA\ AIA band; the correlation with the 171\,\AA\ band was small, so we may conclude that the radio SSS emission comes from a region extending from the chromosphere to the low transition region and does not extend up to the height of formation of the 171\,\AA\ band, which has its peak sensitivity near a temperature of $10^6$\,K.

The exact identification of the SSS with particular structures proved to be extremely difficult, except in the case of patches of old plage and some intense bipolar regions. The rest of the features appear to be due to the merging of several individual structures of the chromospheric network that exist within the instrumental beam. 

Our results are in conformity with the conclusion of previous observations (summarized by \opencite{2011SoPh..273..309S}), as far as the association of radio structures with the chromospheric network is concerned, moreover we cover here a more extended frequency range than previous observations. As for the network itself, it is well known from EUV data that it becomes diffuse in the upper transition region and disappears in the low corona; for example, there is no trace of the network in SDO images in the 171\,\AA\ band. A similar behavior is expected for the radio network and our results are consistent with this. 

In addition to the association of radio SSS with EUV structures, RATAN-600 data provided measurements of the circular polarization and hence of the magnetic field. Our estimates  of 40 to 200 Gauss should be considered as a weighted average, due to the integration over the instrumental beam.

Although the spatial resolution of our data is much inferior to that obtained by modern instruments operating in the EUV, RATAN-600 observations of SSS have certain merits that we would like to emphasize. The most important one is the possibility to estimate the magnetic field, something that is still beyond the capability of EUV instruments. Moreover, the complete spectral coverage of the RATAN-600 which extends over a frequency range with $f_{max}/f_{min}\sim6$, is an extremely usuful input to detailed modelling of the solar atmosphere from the upper chromosphere to the low corona, where the microwave emission is formed. We note, in this context, that very little use has been made of the microwave fine structure data  in multi-component atmospheric models ({\it cf.} \opencite{1983SoPh...85..237C}). Needless is to add that the interpretation of thermal radio emission is devoid of effects such as abundances, excitation and ionization equilibria {\it etc.\/}, that plague the interpretation of EUV data.

The new radio facilities currently under development will certainly add to our understanding of the physics of the network and its extent into the upper layers of the solar atmosphere. In particular, accurate polarization measurements will be possible with the {\it Jansky Very Large Array}, while the {\it Atacama Large Millimeter/submillimeter Array} (ALMA) is expected to extend our vision to the sub-mm range ({\it cf.} \opencite{2006AA...456..697W}). The multi-frequency version of the {\it Siberian Solar Radio Telescope} (SSRT), currently under development (see \opencite{2013PASJ...65S..19K}), is expected to provide two-dimensional spectral information. Last but not least, the {\it Expanded Owens Valley Solar Array} will provide a wide frequency coverage, although with a limited number of baselines and, moreover, it will be a solar dedicated instrument. Unfortunately, the ultimate solar radio instrument, the {\it Frequency Agile Solar Radiotelescope} (FASR), is still far from being realized.
 
The observed fine structure of the quiet-sun in microwaves apparently reflects manifestations of plasma inhomogeneities in the form of supergranulation or chromospheric network. This is indicated by their the characteristic dimensions and life times \cite{2005SoPh..231....1P}. As suggested by many authors, the transformation of the magnetic energy into thermal energy may occur in the chromosphere. The main candidate here could be magnetic reconnection, as the most efficient mechanism both for the release of energy into the corona and the transfer of cool plasma from the chromosphere to the corona (\opencite{2012SSRv..169..181T},  \opencite{2011A&A...532A.112Z}). \inlinecite{2007ApJ...659.1673A} considered a number of arguments that point to the heating processes in the chromosphere. One of the main problems here is the magnetic complexity in transition region, so it is very difficult to trace the magnetic connectivity from coronal loops down to the photosphere (\opencite{2006A&A...460..901J}). 

Future work on fine structures in the chromosphere and low transition region with large optical and ultraviolet telescopes, as well as observations of fine radio structure in microwaves with radio telescopes with big surface area will improve our understanding of their physics and their dynamics. RATAN-600 provides a large data base of such observations, an extended frequency range and the capability of measuring circular polarization. Therefore, it can have an important contribution to their study.

\end{article} 
\end{document}